\documentclass[12pt,preprint]{aastex}
\usepackage{natbib}

\newcommand{\msun}{M\ensuremath{_{\odot}}}
\newcommand{\av}{A\ensuremath{_{V}}}
\newcommand{\rhooph}{$\rho$ Oph}
\newcommand{\teff}{T$_{\mathrm{eff}}$}
\newcommand{\ks}{K\ensuremath{_{\mathrm{s}}}}

\shorttitle{SONYC: \rhooph}
\shortauthors{Geers et al.}
\slugcomment{Accepted for publication in the Astrophysical Journal 10/25/2010}

\begin{document}

\title{Substellar Objects in Nearby Young Clusters (SONYC) II: \\
The Brown Dwarf Population of $\rho$ Ophiuchi\thanks{Based on data collected at Subaru Telescope, which is operated by the National Astronomical Observatory of Japan.}}

\author{Vincent Geers\altaffilmark{1,2}, Alexander Scholz\altaffilmark{3}, Ray Jayawardhana\altaffilmark{1,**}, Eve Lee\altaffilmark{1}, David Lafreni\`{e}re\altaffilmark{1,4}, Motohide Tamura\altaffilmark{5}}

\email{vcgeers@astro.utoronto.ca}

\altaffiltext{1}{Department of Astronomy \& Astrophysics, University of Toronto, 50 St. George Street, Toronto, ON M5S 3H4, Canada}
\altaffiltext{2}{Institute for Astronomy, ETH Zurich, Wolfgang-Pauli-strasse 27, 8093 Zurich, Switzerland}
\altaffiltext{3}{School of Cosmic Physics, Dublin Institute for Advanced Studies, 31 Fitzwilliam Place, Dublin 2, Ireland}
\altaffiltext{4}{D\'{e}partement de Physique, Universit\'{e} de Montr\'{e}al, C.P. 6128 Succ. Centre-Ville, Montr\'{e}al, QC H3C 3J7, Canada}
\altaffiltext{5}{National Astronomical Observatory, Osawa 2-21-2, Mitaka, Tokyo 181, Japan}
\altaffiltext{**}{Principal Investigator of SONYC}

\begin{abstract}
SONYC -- {\it Substellar Objects in Nearby Young Clusters} -- is a survey program to investigate the frequency and properties of brown dwarfs down to masses below the Deuterium burning limit in nearby star forming regions. In this second paper, we present results on the $\sim$1\,Myr old cluster $\rho$ Ophiuchi, combining our own deep optical and near-infrared imaging using Subaru with photometry from the 2-Micron All Sky Survey and the {\it Spitzer Space Telescope}. Of the candidates selected from iJ\ks\ photometry, we have confirmed three -- including a new brown dwarf with a mass close to the Deuterium limit -- as likely cluster members through low-resolution infrared spectroscopy. We also identify 27 sub-stellar candidates with mid-infrared excess consistent with disk emission, of which 16 are new and 11 are previously spectroscopically confirmed brown dwarfs. The high and variable extinction makes it difficult to obtain the complete sub-stellar population in this region. However, current data suggest that its ratio of low-mass stars to brown dwarfs in similar to those reported for several other clusters, though higher than what was found for NGC 1333 in \citet{scholz09}.
\end{abstract}


\section{Introduction}
\label{sec:intro}
Understanding the origin of the stellar initial mass function (IMF) is one of the major topics in astrophysics. The low-mass end of the IMF, in particular, has been subject of numerous observational and theoretical studies over the past decade \citep[see][]{bonnell07}. 

SONYC -- Substellar Objects in Nearby Young Clusters -- is an ongoing project to provide a complete census of the brown dwarf (BD) and planetary mass object population in nearby young clusters, and to establish the frequency of substellar mass objects as a function of cluster environment. The resulting catalog of substellar mass candidates will provide the basis for detailed characterization of their physical properties (disks, binarity,  atmospheres, accretion, activity). The survey makes use of  extremely deep optical and near-infrared imaging, combined with the 2MASS and Spitzer photometry catalogs, and follow-up spectroscopy using 8-m class telescopes, aiming to detect the photosphere of the objects, which is an essential prerequisite for an unbiased survey. Our observations are designed to reach limiting masses of 0.003 \msun, well below the Deuterium burning limit at 0.015 \msun, and in each region, we aim to cover at least $\sim 1000$\,arcmin$^2$.

In our first paper \citep{scholz09}, we presented the survey of the NGC 1333 cluster in Perseus, which was surveyed complete down to 0.004-0.008 \msun\ for \av\ up to 5--10 mag. We identified 12 new substellar members in this cluster based on photometry and spectroscopy, and found a tentative minimum mass cut off to the IMF at 0.012-0.020 \msun.

In this second paper, we report on the survey of the \rhooph\ star forming cloud complex. \rhooph\ is one of the closest (d = 125$\pm$25 pc, \citealp{de-geus89}) regions of active star formation. In contrast with the first SONYC field, NGC 1333, the \rhooph\ cluster is not as compact and exhibits extremely high and variable levels of extinction. The main cloud, L1688, is a dense molecular core, with visual extinction up to 50--100 mag \citep{wilking83}, hosting an embedded infrared cluster of around 200 stars, inferred to have a median age of 0.3 Myr, and surrounded by multiple clusters of young stars with a median age of 2.1 Myr \citep[and references therein]{wilking05}. 

To date, a population of 17 BD candidates in \rhooph, with spectral M6 or later, have been spectroscopically confirmed by \citet{wilking99}, \citet{luhman99}, and \citet{cushing00}. An ISOCAM survey of very young stars in \rhooph\ was presented in \citet{bontemps01}, from which a selection of candidate BDs with disks was discussed in \citet{natta02}. For a number of BDs, \citet{jayawardhana02} searched for accretion signatures with high resolution optical spectroscopy, and \citet{jayawardhana03b} found a disk fraction of $\sim$67\% for \rhooph\ from L$'$-band (3.8 \micron) excess. A survey of accretion rate measurements for the full ISOCAM survey was presented in \citet{natta06}. For a recent summary of the star formation and brown dwarf population in \rhooph, see \citet{wilking08}. A few brown dwarfs were found and spectroscopically confirmed at about 1 degree away from the center of the L1688 cloud \citep{jayawardhana06b,allers07}.

This paper is structured as follows. Observations and data reduction of imaging and spectroscopy are described in Sect.\ \ref{sec:obsred}. The photometric selection of very low mass objects is described in Sect.\ \ref{sec:selection}, and spectroscopy is presented in Sect.\ \ref{sec:mosspec}. The results are discussed in Sect.\ \ref{sec:discussion}, and conclusions are presented in Sect.\ \ref{sec:conclusions}.

\section{Multi-band photometry}
\label{sec:obsred}

\subsection{Optical imaging}
\label{ssec:optphot}

We obtained optical images with the Subaru Prime Focus Camera (Suprime-Cam) wide field imager \citep{miyazaki02} in the  SDSS i' filter on August 2, 2008, with a typical range in seeing of 0.50-0.60\arcsec. Suprime-Cam is a mosaic camera utilizing 10 CCDs arranged in a 5 $\times$ 2 pattern, giving a total field of view of  $34'\times27'$, with a spatial resolution of 0.20$''$ per pixel. The \rhooph\ cluster was observed in a single field with 15 individual exposures of 120 sec in a 10-point dither pattern. In total this gives an integration time of 1800 sec.

We performed image reduction separately for each individual chip, using the Suprime-Cam reduction software package {\em SDFRED} \citep{yagi02,ouchi04}. This includes overscan subtraction, flat-fielding, distortion correction, sky subtraction, bad pixel and AG probe masking, and finally image combination. The 2MASS point source catalog \citep{cutri03} was used as a reference for the calibration of the world coordinate system, using the {\em msctpeak} routine from the IRAF package {\em MSCRED}\footnote{IRAF is distributed by the National Optical Astronomy Observatories, which are operated by the Association of Universities for Research in Astronomy, Inc., under cooperative agreement with the National Science Foundation.}. The typical fitting residuals were of order 0.15\arcsec.

The objects were identified using the Source Extractor ({\em SExtractor}) software package \citep{bertin96}. We require an object to have at least 5 pixels with flux above the 3\,$\sigma$ detection limit to be extracted. The automatic aperture fitting routine in {\em SExtractor} was used to calculate the flux of each source. We chose conservative rejection criteria, to minimize the number of spurious detections in the photometry database, including the rejection of objects within 25 pixels of the edge of the image, elongated objects ($a/b>1.2$), and saturated objects. The results from the detection and rejection algorithm were verified by visual inspection, to minimize the number of spurious detections and missed point-sources. The final optical catalog has 5791 objects. 

\begin{figure}
\center
\includegraphics[width=\columnwidth]{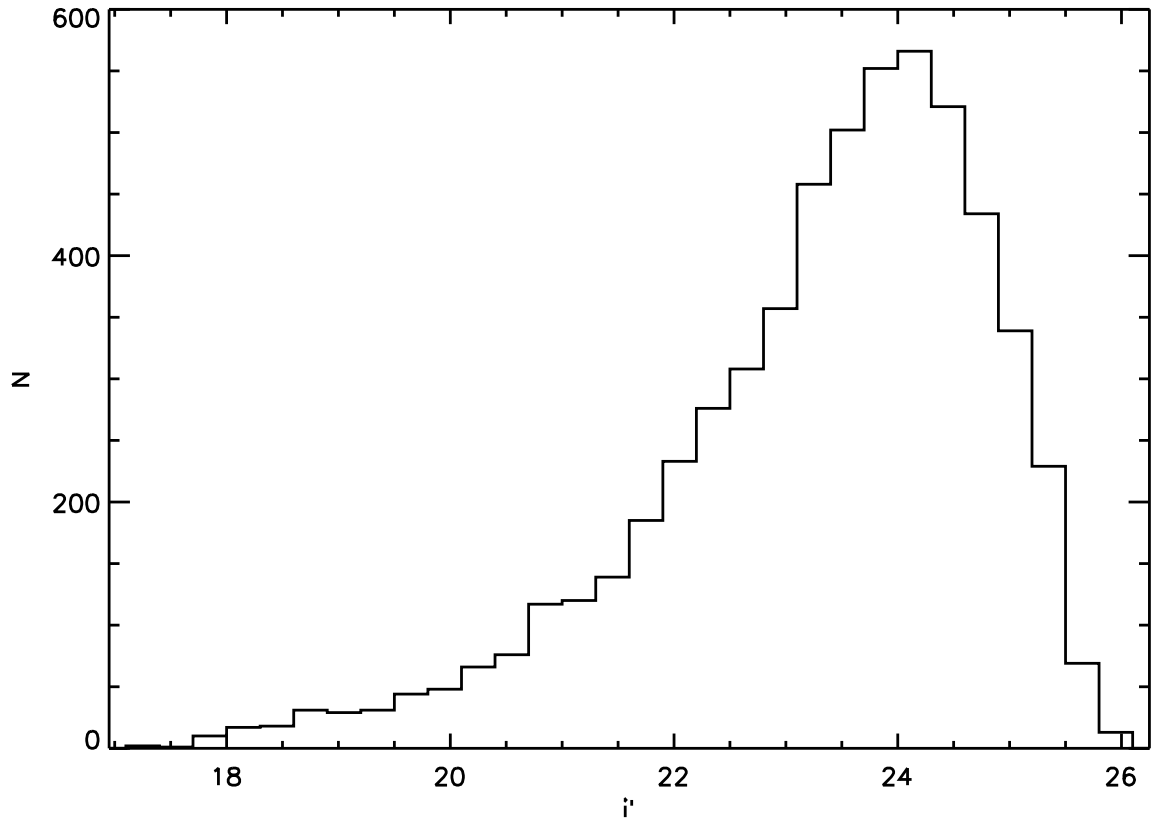} 
\caption{i'-band histogram of the objects in our photometric catalog with i'-band data. The peak at 24.15\,mag indicates the completeness limit of the survey. The faintest objects in the survey are found at $i>26$\,mag.
\label{fig:icomp}}
\end{figure}

The chip-to-chip zeropoint offsets were derived from the median fluxes of domeflat images. The absolute zeropoint for the mosaic was derived from observations of the SA112 standard field, which contains SDSS secondary standards \citep{smith02}. These standard observations were taken at the same airmass and with the telescope defocused to avoid saturation. Four standard stars were available in this field and photometry was obtained using a fixed 30-pixel radius aperture extraction. For the i'-band, we derive a mean zeropoint of $32.588$~$\pm$~$0.076$ mag, and a completeness limit of 24.15 $\pm$ 0.30 mag, see Fig.\ \ref{fig:icomp}. 

\begin{figure}
\center
\includegraphics[width=\columnwidth]{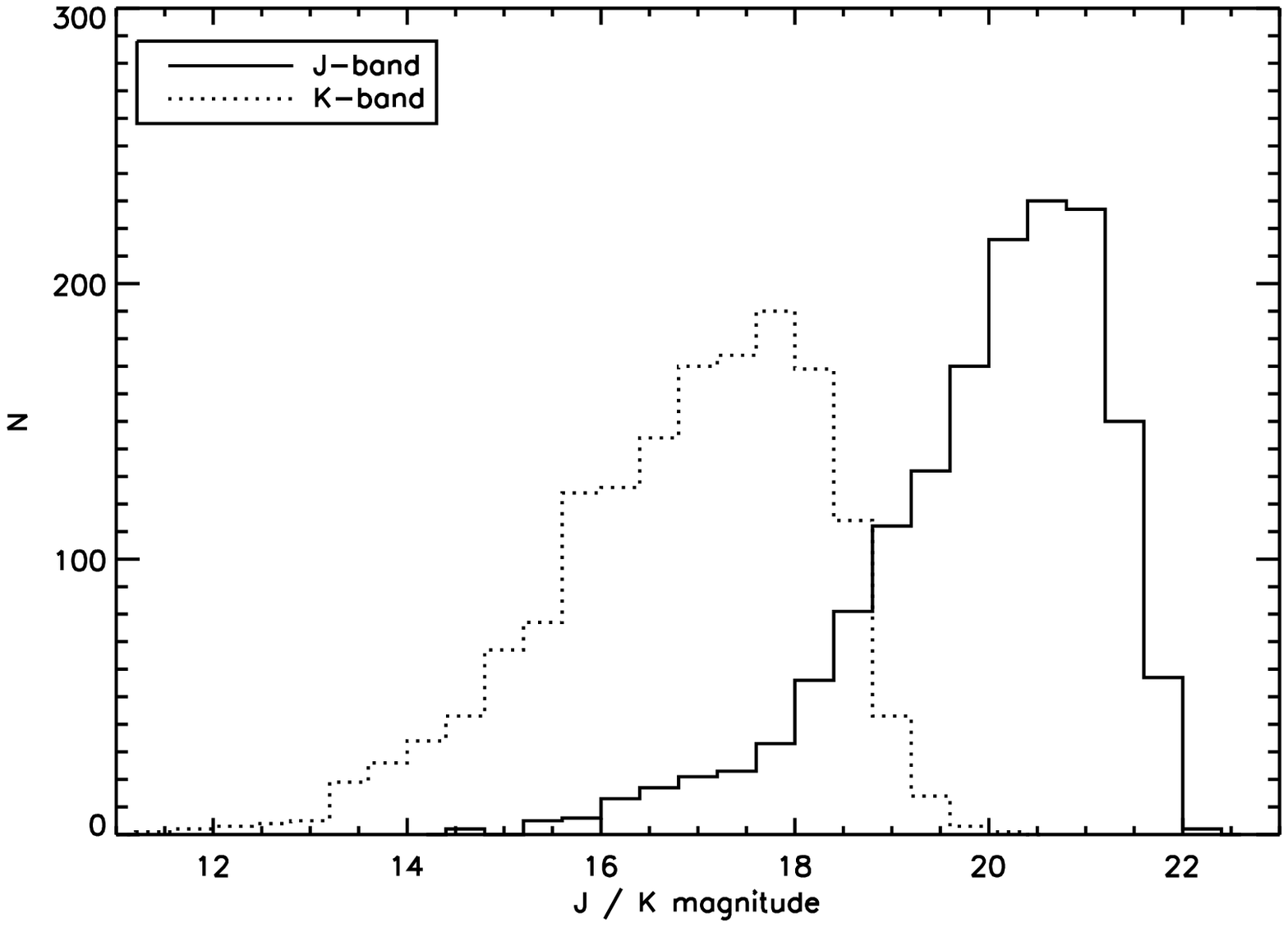} 
\caption{J-band and \ks-band histogram of the objects in our photometric catalog with J- and \ks-band data. The peaks at 20.6 and 17.8\,mag indicates the completeness limit of the survey in J- and \ks-band respectively. 
\label{fig:jkcomp}}
\end{figure}

\subsection{Near-infrared imaging}
\label{ssec:nirphot}
The Multi-Object Infrared Camera and Spectrograph (MOIRCS) \citep{suzuki08} mounted on the Subaru telescope was used to observe \rhooph\ in the J- (1.26 \micron) and \ks-band (2.14 \micron). The data was obtained June 22 to June 24, 2007. MOIRCS uses two detector arrays which provide a field of view of 4$'$ by 7$'$ in each pointing. We observed 24 MOIRCS fields with an overlap of 30$''$ between adjacent fields, encompassing the embedded subclusters \rhooph\ A, B, E, and F in the L1688 cloud \citep{loren90}. The total imaging area coverage was 31.5$'$ by 26$'$. Each field was covered in a six-point dither pattern with individual exposure times of 100 s in J-, and 50 s in \ks-band. This yields a total integration time of 600 s in J- and 300 s in \ks-band. Significant contamination by a bright nearby source  was present in 15 individual images. 

The reduction was carried out with the SIMPLE-MOIRCS package\footnote{http://subarutelescope.org/Observing/Instruments/MOIRCS/imag\_information.html}. First, we corrected for the offset in sensitivity between the two detector arrays. The individual frames are then flatfield corrected for flatfields and co-added by median. For the source detection we used the SExtractor, requiring at least 5 pixels with flux above the 3.5\,$\sigma$ detection limit. Saturated and extended sources were rejected from further consideration. We calibrated the coordinate system against the 2MASS point source catalog \citep{skrutskie06} for J-band and \ks-band separately, with typical fitting residuals of 0.11\arcsec. Duplicate sources from adjacent fields were removed. We cross-correlated the sources from the J-band catalog with the \ks-band catalog, requiring a separation of $< 2''$ to be a match. Sources that are only detected in one band, either J or \ks, were rejected. The final near-infrared catalog contains 1571 objects.

The fluxes for these sources were extracted using a fixed 6 pixel radius aperture for all fields. To correct for the variable seeing, we derived an aperture correction factor for each image and applied it to the fluxes. Based on the 2MASS point source catalog, we derived the absolute zero point for each individual field. For 19 fields no 2MASS sources were covered; for these fields a zero point is derived based on the average of zero points in other fields at similar airmass. The completeness limit of the observations was found to be 20.6$\pm$0.30 in J-band and 17.8$\pm$0.30 in \ks-band, see Fig.\ \ref{fig:jkcomp}.

\subsection{Cross correlation of optical and NIR catalogs}
The sources from the i'-band catalog were cross-correlated by coordinates with the sources from the J\ks\ catalog, requiring a separation of $< 2''$ to be a match. The resulting final catalog of sources with detections in all three bands (hereafter ``iJ\ks\ catalog'') contains 504 sources.

\begin{figure}
\centering
\includegraphics[width=0.7\columnwidth]{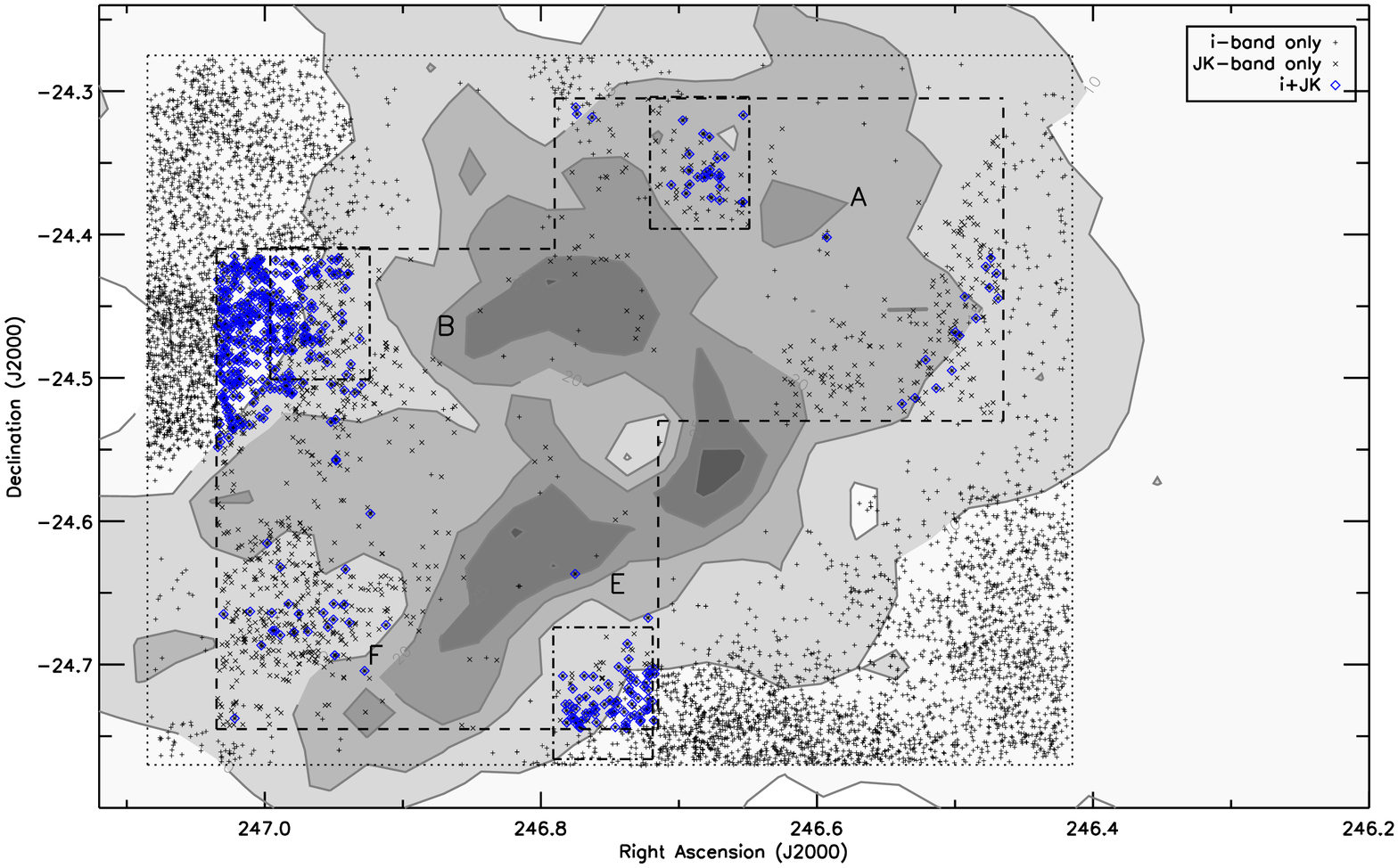}
\includegraphics[width=0.7\columnwidth]{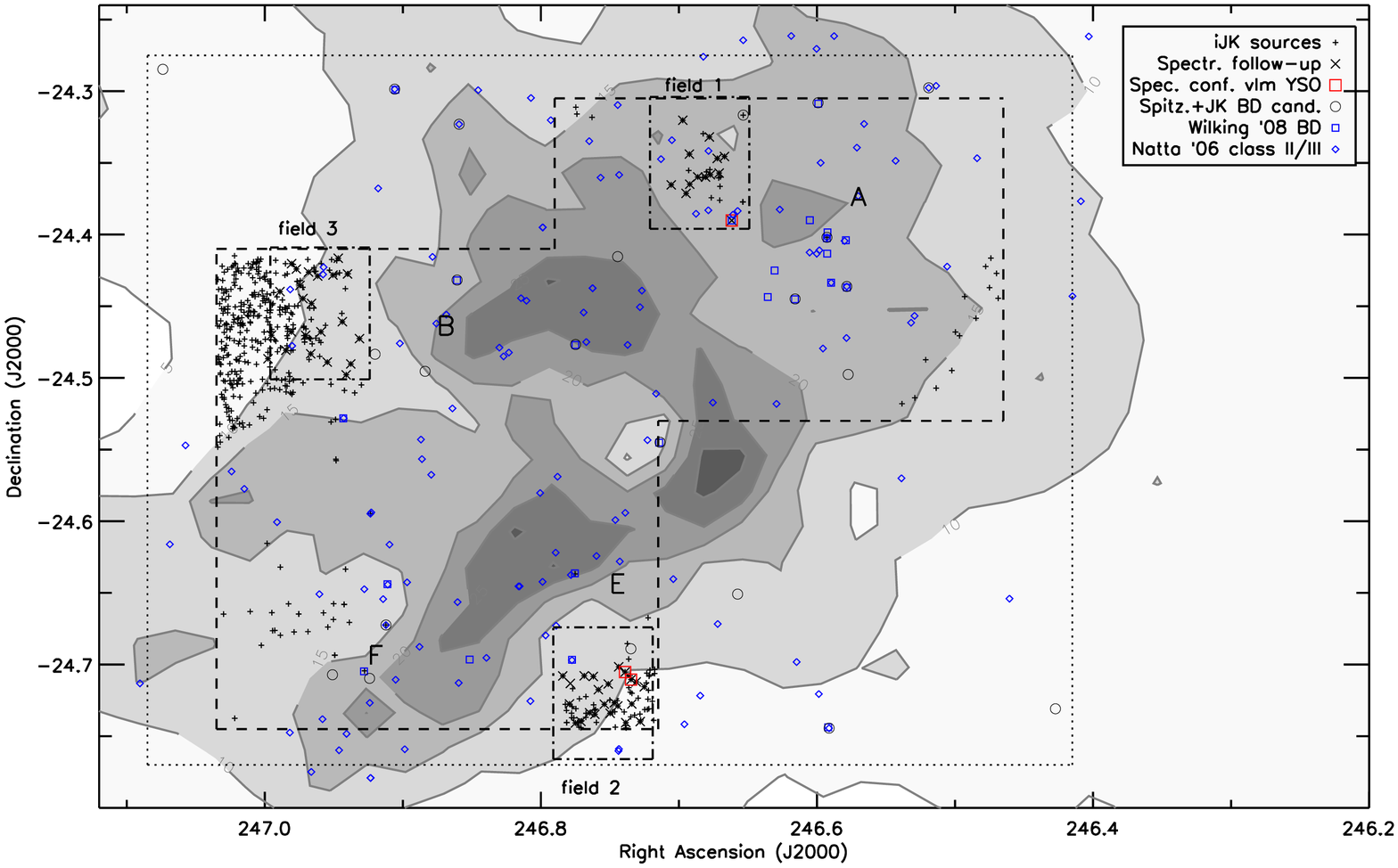}
\caption{Spatial distribution of sources in \rhooph. Contours are \av\ = 5, 10, 15, 20, 25, 30, as derived from 2MASS by the COMPLETE project. Dotted line: i-band imaging coverage; dashed line: J and \ks-band imaging coverage; dash-dot line: MOIRCS spectroscopy fields. The locations of \rhooph\ subclusters, as defined by the peaks of the DCO$^+$ dense cores from \citet{loren90}, are indicated with letters. {\bf Top panel}: i-band indicated +, J and \ks-band indicated with x, iJ\ks\ catalog sources indicated with diamonds. {\bf Bottom panel}: iJ\ks\ catalog sources indicated with +; iJ\ks\ catalog selected BD candidates with follow-up MOIRCS spectroscopy indicated with x (Sect.\ \ref{sec:mosspec}); low mass YSOs spectroscopically confirmed in this paper (Sect.\ \ref{ssec:specana}) indicated with red large squares. Candidate BDs selected from J\ks\ + Spitzer photometry indicated with circles. Previously known BDs from \citet{wilking08} and class II / III sources from \citet{natta06} are indicated with small blue squares and diamonds respectively. }
\label{fig:spatial}
\end{figure}

The spatial distribution of these 504 sources is shown in Fig.\ \ref{fig:spatial} in comparison with the visual extinction map given by the COMPLETE project\footnote{http://www.cfa.harvard.edu/COMPLETE}. As seen in the figure, we have chosen the pointings of our observations to cover the high extinction contours of the L1688 core; however, parts of the region surveyed by Suprime-Cam were not covered in MOIRCS.

The majority of the iJ\ks\ catalog objects are located outside the region with the highest \av, at a typical \av\ range of 5-15.
At higher \av, toward the center of \rhooph, the lack of sources corresponds to a lack of i'-band detections due to the high \av. At lower \av, many objects are saturated in J\ks. As shown in Fig.\ \ref{fig:spatial}, the majority of the BD candidates are found in several distinct clumps at regions with low to moderate cloud extinction, \av\ = 5--15, within the region covered by both the Suprime-Cam and MOIRCS imaging. Three of these regions are selected for multi-object follow-up spectroscopy, see Sect.\ \ref{sec:mosspec}, which showed that this apparent clustering is dominated by background objects. Thus, the distribution of the combined iJ\ks\ catalog sources traces substructure in the extinction, and not in the distribution of YSOs.

We convert the dynamic range and completeness limit of the i'-band and J-band survey separately to mass sensitivity limits, by comparing with model evolutionary tracks DUSTY00 \citep{allard01} and COND03 \citep{baraffe03}, described further in Sect.\ \ref{ssec:ijk_candsel} and Figs.\ \ref{fig:cmd_i_ij} and \ref{fig:cmd_j_jk_ijk_bdcand}. 

The majority of i'-band magnitudes are between 18 and 25.5 mag, with the completeness limit at 24.15 mag. \rhooph\ has significant and strongly variable extinction, with a typical minimum \av\ of 5 mag, ranging up to 15--30 mag near the center of the cluster, and upwards of 30 mag in the very center of the cluster \citep{ridge06}. For an \av\ of 5, corresponding to A$_{\mathrm{i}}$ = 3.45 \citep{mathis90}, the i'-band range of 18--25.5 mag corresponds to masses ranging between 0.002--0.068 \msun, with the completeness limit in i'-band corresponding to a mass ranging between 0.004 to 0.005 \msun\, based on the 1 Myr COND03 and DUSTY00 models. 
At a moderate \av\ = 15, the i'-band survey range corresponds to masses from 0.047 to more than 0.1 \msun (outside model range), with the completeness limit corresponding to a mass sensitivity limit of 0.09 to 0.1 \msun. 

The J-band survey was sensitive to magnitudes between 15 and 21.3, with a completeness limit of 20.6 mag. For \av\ = 5 (A$_{\mathrm{J}}$ = 3, \citealt{mathis90}) the 15--21.3 magnitude range corresponds to masses in the range of 0.001--0.034 \msun, with the completeness limit corresponds to masses of $\sim 0.001$ -- 0.003 \msun. 
For an \av\ of 15, the survey range corresponds to masses of 0.005--0.1 \msun, with the completeness limit corresponding to masses of 0.006-0.007 \msun. 

In summary, the sensitivity of the i'-band and J-band survey were such that they were complete down to below the substellar regime, for low to moderate \av, while sensitivity in the upper mass range was different between i'-band and J-band, due to J-band saturating at lower masses. However, particularly at large \av, the survey was still not deep enough in i'-band to probe down into the substellar regime, and below.

\subsection{Cross correlation of optical and 2MASS NIR catalogs}
\label{ssec:2mass_optical}
To increase the dynamic range of the survey in the near-infrared, we retrieved additional J\ks\ photometry from the 2MASS point-source catalog \citep{skrutskie06}. We selected 2MASS sources located within the spatial coverage of our i'-band survey. The objects were further required to not be extended with semi-major axis $>$ 10\arcsec, nor fall within the elliptical profile of such an extended source (xflag = 0), and not to be associated with known solar system objects (a-flag = 0) and detection quality flags in J and \ks-band of `A'. The resulting 2MASS catalog was cross-correlated by coordinates with the i'-band catalog sources. We require a separation of $< 2''$ to be a match. The final i' + 2MASS J-, and \ks-band catalog contains 264 sources. The majority of these sources were saturated in the MOIRCS data, as illustrated by Fig.\ \ref{fig:cmd_i_ij}.

\subsection{Spitzer photometry}
\label{ssec:spitzerphot}
We retrieved the Spitzer IRAC and MIPS photometry of \rhooph\ from the Legacy Program data archive available at the Spitzer Science Center, using the ``High reliability catalog'' (HREL) created by the ``Cores to Disks'' (\emph{c2d}) Legacy team, made available at the SSC website\footnote{http://data.spitzer.caltech.edu/popular/c2d/20071101\_enhanced\_v1/oph/catalogs/}. The photometry from this catalog is used in Sect.\ \ref{ssec:jk_spitzer_candidates} to search for candidate substellar objects with disks. 

The Spitzer catalog was cross-correlated with the MOIRCS J+\ks-band catalog as well as the 2MASS catalog (see selection in Sect.\ \ref{ssec:2mass_optical}). Here we require a separation of $< 5''$ for a match, and the final catalog is restricted to the spatial area of the Suprime-Cam i'-band coverage (246.415 $\le$ RA $\le$ 247.085, -24.77 $\le$ DEC $\le$ -24.275). The resulting catalog contains 952 matches between MOIRCS and Spitzer, and 1166 matches between Spitzer and 2MASS.

\section{Selection of candidate low mass sources in \rhooph}
\label{sec:selection}

\subsection{iJ\ks\ candidate selection}
\label{ssec:ijk_candsel}
\begin{figure}
\includegraphics[width=\columnwidth]{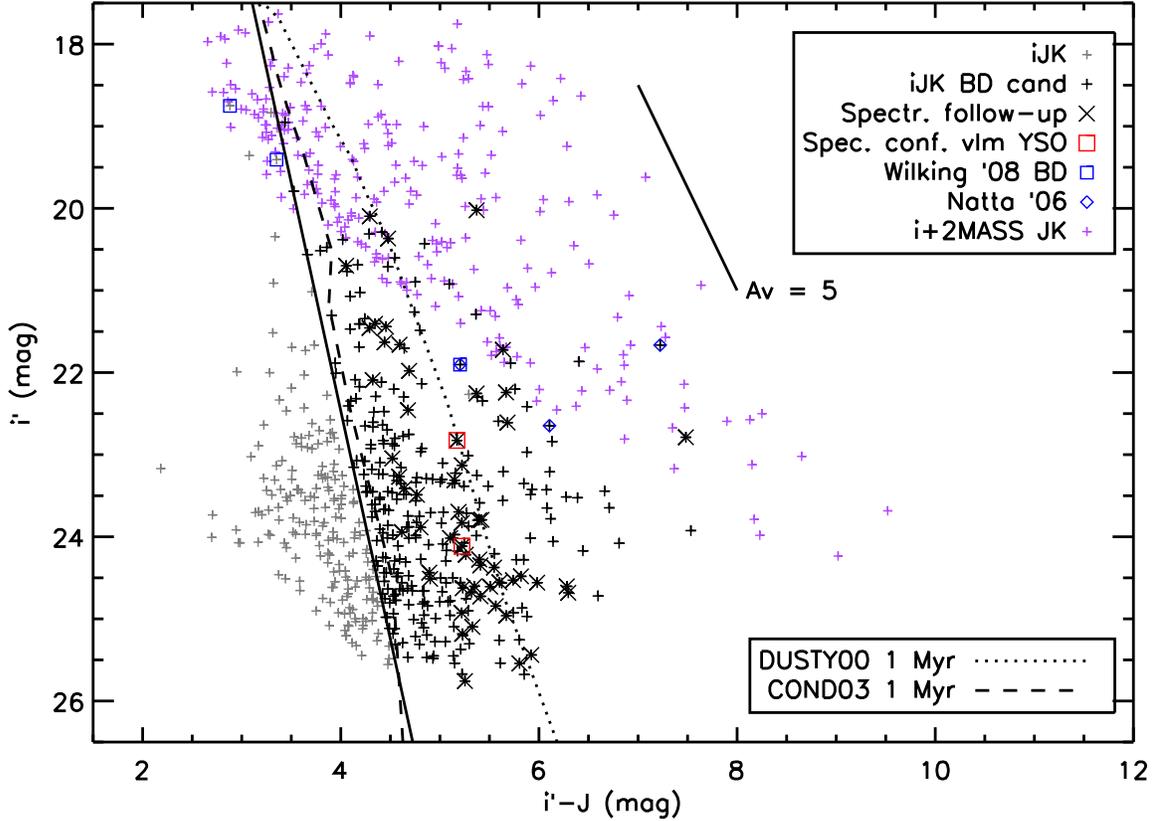}
\caption{Colour-magnitude diagram in (i', i'--J). Sources from the Suprime-Cam and MOIRCS iJ\ks\ catalog are indicated with light ``+'' symbols, while BD candidates based on iJ\ks\ are indicated in dark ``+'' symbols. Sources from the Suprime-Cam and 2MASS catalogs are indicated with purple ``+'' symbols. Spectroscopy follow-up targets are indicated with x, two confirmed very low mass YSOs with large red squares.  Previously known BDs from \citet{wilking08} are indicated with small blue squares, class II / III sources from \citet{natta06} are indicated with small blue diamonds. Atmosphere model isochrones for age 1 Myr are overplotted for DUSTY00 \citep{allard01} and COND03 \citep{baraffe03}. The solid line denotes the BD candidate selection cut-off.}
\label{fig:cmd_i_ij}
\end{figure}

We initially selected low mass and BD candidates based solely on the (i', i'--J) colour-magnitude diagram. Fig.\ \ref{fig:cmd_i_ij} shows the colour-magnitude diagram constructed from the Suprime-Cam i'-band catalog and both the MOIRCS and the 2MASS J-, + \ks-band catalog. For comparison, the DUSTY00 \citep{allard01} and COND03 \citep{baraffe03}  atmosphere model evolutionary tracks for an age of 1 Myr are overplotted, adjusted to the distance of \rhooph\ of 125 pc. We converted the Cousins I-band to the Sloan i'-band using the transformation given by \citet{jordi06}. The BD candidate selection cut-off in (i'--J) is shown in Fig.\ \ref{fig:cmd_i_ij} and was determined as a linear approximation of the 1 Myr COND03 track. Any object redder than this line is considered a candidate.
This selection gives us 309 BD candidates based on our iJ\ks\ catalog and an additional 228 sources based on the Suprime-Cam i'-band + 2MASS J\ks\ catalog. We obtained follow-up spectroscopy for a selection of the iJ\ks\ catalog based BD candidates, as presented in Sect.\ \ref{sec:mosspec}.

The intrinsic J--\ks\ color of late-type objects is mostly independent of spectral type and luminosity and can be used to estimate the extinction for suspected cluster members. We adopt here an intrinsic color of J--\ks\ = 1.0 mag, which is consistent, within $\pm\ 0.2$ mag, with the empirical values for main sequence dwarfs and giants with spectral type M \citep{bessell88}, with the predictions from the evolutionary tracks COND03 (for 0.003--0.1 \msun), BCAH98 (for 0.03--0.5 \msun), and DUSTY00 (for 0.006--0.1 \msun), and with the colors of BDs in the 3 Myr old $\sigma$ Ori star forming region down to 0.013 \msun\ \citep{caballero07}. For young objects with masses below the Deuterium burning limit, there are indications that the J--\ks\ color increases to 1.5 \citep{Lodieu06, caballero07}; similarly the presence of a circumstellar disk can increase the J--\ks\  color. 

\begin{figure}
\includegraphics[width=\columnwidth]{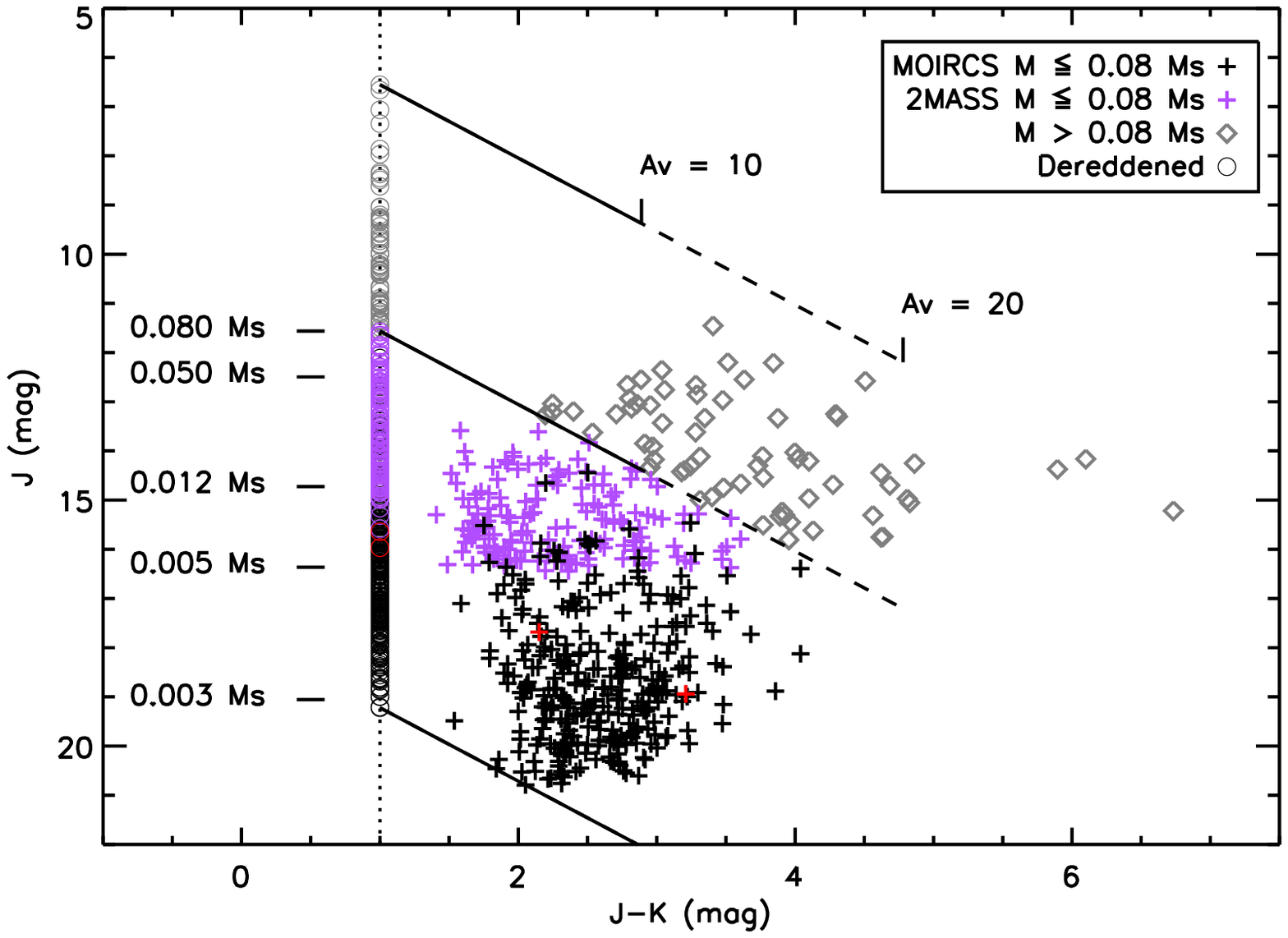}
\caption{Colour-magnitude diagram in (J, J--\ks), constructed from the MOIRCS and 2MASS photometry, for the BD candidates selected from the iJ\ks\ catalog and the i-band + 2MASS catalog. Solid and dashed lines indicate visual extinction of 10 and 20 magnitude from the assumed intrinsic J--\ks\ of 1 (vertical dotted line). ``+'' and diamond symbols indicate the MOIRCS and 2MASS photometric points, circles are dereddened in J. The two very low mass candidates confirmed by MOIRCS spectroscopy are included with red symbols.}
\label{fig:cmd_j_jk_ijk_bdcand}
\end{figure}

To estimate the extinction we use the reddening law from \citet{mathis90} with $R_{V} = 4$. Fig.\ \ref{fig:cmd_j_jk_ijk_bdcand} shows the (J, J--\ks) colour-magnitude diagram for the sample of 309+228 i'-band + MOIRCS and 2MASS catalog derived BD candidates. The reddening path is indicated in solid and dashed lines, and the intrinsic color indicated with a vertical dotted line. The 1Myr COND03 evolutionary track is used for the mass estimate based on the de-reddened J-band magnitude. We obtain visual extinctions of \av\ = 3 -- 16 for the MOIRCS sources and \av\ = 2 -- 34 for the 2MASS sources. 

The main errors in the extinction estimates are the uncertainties in the estimate of the intrinsic colors (accurate to within 0.2 mag in J--\ks) and the possible excess emission due to disks and accretion ($<1$ mag in J--\ks, see \citealt{meyer97}), particularly for the sample selected based on Spitzer MIR excess in Sect.\ \ref{ssec:jk_spitzer_candidates}.  \citet{wilking08} lists 316 L1688 association members known in January 2008, out of which 152 show infrared excess indicative of the presence of a disk. (indicated with `B2ex', `B4ex', and `Mex' in their Table 1). This illustrates that a substantial fraction of the L1688 members may show enhanced \ks-band flux due to excess emission from the dust in the disk. We conservatively estimate that our typical uncertainty in \av\ is in the range of $\pm 1$ mag for most sources, but may be up to 5 mag if disk excess is present. We note that brown dwarfs with disks typically have only little colour excess in the near-infrared.

The estimated range of visual extinction is in good agreement with the \rhooph\ extinction map as derived from 2MASS imaging observations by the COMPLETE team. 

\subsection{J\ks\ + Spitzer candidate selection}
\label{ssec:jk_spitzer_candidates}
\begin{figure}
\includegraphics[width=\columnwidth]{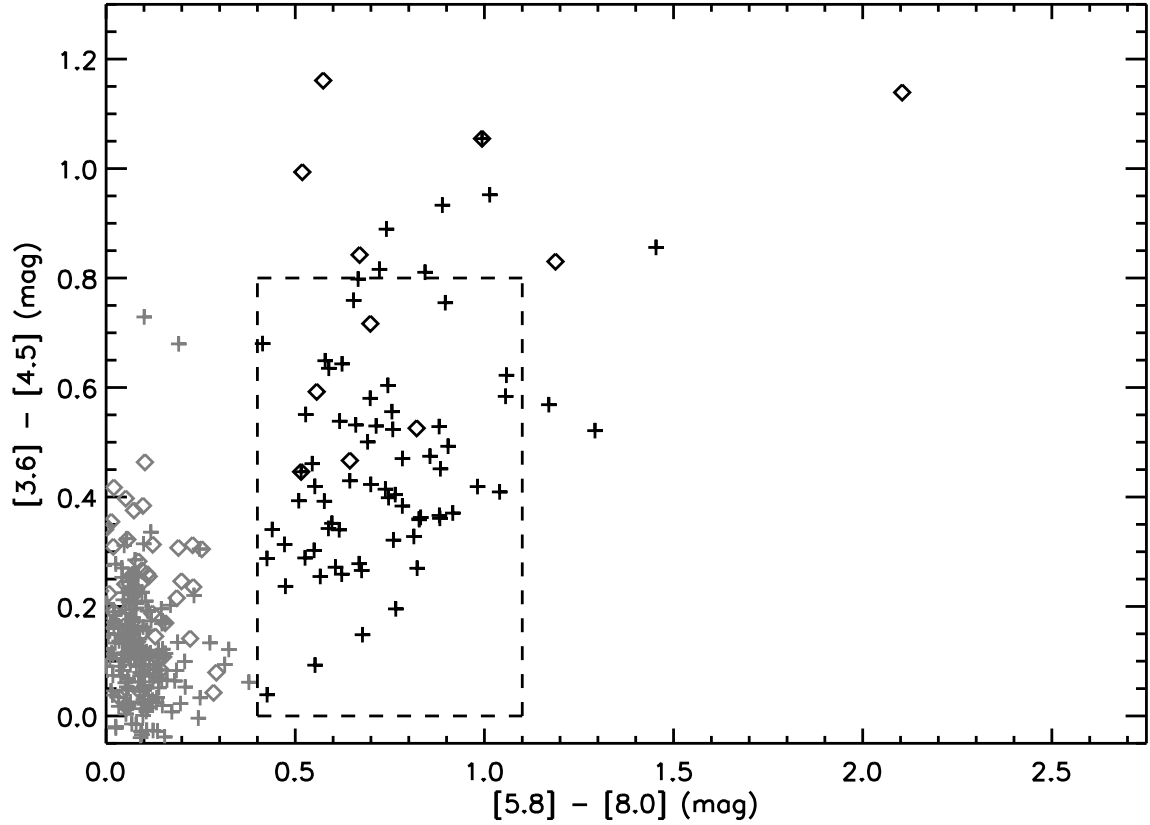}
\caption{Colour-colour diagram in (3.6--4.5, 5.8--8.0), constructed from Spitzer IRAC photometry, for 2MASS (crosses) and MOIRCS (diamonds) sources with a Spitzer match. The dashed line denotes the area where class II objects are located, based on \citet{allen04}. All sources selected as having significant NIR excess are indicated with in black.}
\label{fig:ccd_irac}
\end{figure}

As the i'-band observations were not complete in the central parts of L1688 with high \av, we selected a second sample of candidate substellar mass objects with disks based on the mid-infrared photometry from the Spitzer Telescope, restricted to the spatial area of the i'-band coverage (see Sect.\ \ref{ssec:spitzerphot}). Fig.\ \ref{fig:ccd_irac} shows a colour-colour diagram of the 4 IRAC bands for the MOIRCS and 2MASS sources. For 11 MOIRCS sources, and 72 2MASS sources, the IRAC colours are consistent or redder than sources with a circumstellar disk. The presence of a circumstellar disk is taken as a confirmation of youth and membership of the \rhooph\ cluster.

\begin{figure}
\includegraphics[width=\columnwidth]{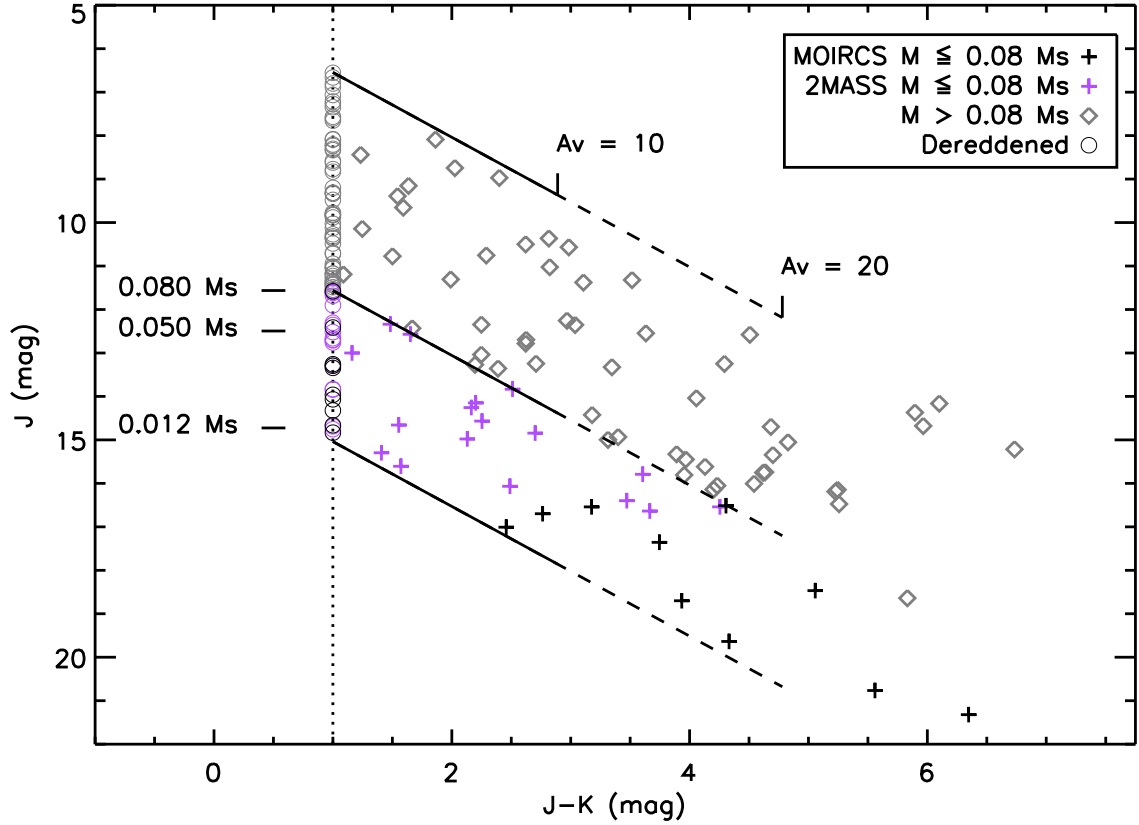}
\caption{Colour-magnitude diagram in (J, J--\ks), constructed from the MOIRCS and 2MASS photometry, for the sources with Spitzer NIR excess. Solid and dashed lines indicate visual extinction of 10 and 20 magnitude from the assumed intrinsic J--\ks\ of 1. ``+'' and diamond symbols indicate the MOIRCS and 2MASS photometric points, circles are dereddened in J.}
\label{fig:cmd_j_jk_cand}
\end{figure}

Fig.\ \ref{fig:cmd_j_jk_cand} shows the (J, J--\ks) colour-magnitude diagram for the sample of 11 MOIRCS and 72 2MASS sources with excesses in the Spitzer data. 
Based on the CMD and the DUSTY00 atmospheric model, we find 10 MOIRCS BD candidates and 17 2MASS BD candidates with M $<$ 0.08 \msun, assuming an intrinsic brown dwarf photosphere J--\ks\ color of 1 magnitude. These 10 + 17 candidates are listed in Table \ref{tbl:subst_disks}. Among this selection of 27 BD candidates, 16 are new, while 11 are previously spectroscopically confirmed BDs. If we assume an additional 1 magnitude J--\ks\ excess due to the presence of a circumstellar disk (intrinsic J--\ks\ = 2), then the number of candidates increases to 11 MOIRCS and 39 2MASS sources.

\clearpage
\begin{deluxetable}{lllllllllllll}
\rotate
\tabletypesize{\scriptsize}
\tablecolumns{13}
\tablecaption{Likely substellar members with disks within i'-band surveyed area within \rhooph\ from \emph{Spitzer} + J\ks\ photometry.\label{tbl:subst_disks}}
\tablewidth{0pt}
\tablehead{
\colhead{\#} & \colhead{RA} & \colhead{Dec} & \colhead{J} & \colhead{\ks} & \colhead{IRAC1} & \colhead{IRAC2} & \colhead{IRAC3} & \colhead{IRAC4} & \colhead{\av} & \colhead{J\ks\ phot.} & \colhead{Identifiers} & \colhead{References\tablenotemark{a}}\\ 
\colhead{} & \colhead{(J2000)} & \colhead{(J2000)} & \colhead{(mag)} & \colhead{(mag)} & \colhead{(mJy)} & \colhead{(mJy)} & \colhead{(mJy)} & \colhead{(mJy)} & \colhead{(mag)} & \colhead{} & \colhead{} & \colhead{BD candidacy}
}
\startdata
  1 &  16 26 18.58 &   -24 29 51.4 & 17.36 &  13.61 &   3.93 &   4.08 &   4.58 &   5.44 &  14.5  & MOIRCS & BKLT J162618-242951 & -\\ 
  2 &  16 26 22.28 &   -24 24 07.1 & 16.70 &  13.94 &   3.10 &   4.26 &   5.53 &   9.21 &   9.3  & MOIRCS & GY 11 & WGM99,CTK00,N02\\
  3 &  16 26 36.84 &   -24 19 00.1 & 17.01 &  14.55 &   1.10 &   2.05 &   3.70 &   3.50 &   7.7  & MOIRCS & BKLT J162636-241902 & -\\ 
  4 &  16 26 56.37 &   -24 41 20.5 & 18.70 &  14.77 &   1.01 &   0.99 &   1.06 &   1.07 &  15.5  & MOIRCS & - & - \\
  5 &  16 26 58.66 &   -24 24 55.6 & 19.64 &  15.30 &   3.72 &   5.17 &   5.65 &   5.84 &  17.6  & MOIRCS & AOC J162658.65-242455.5 & -\\
  6 &  16 26 59.04 &   -24 35 56.9 & 16.51 &  12.21 &  14.20 &  13.70 &  11.60 &  10.40 &  17.5  & MOIRCS & GY 172 & - \\  
  7 &  16 27 32.15 &   -24 29 43.6 &   18.47 &  13.41 &   8.28 &   9.14 &   9.16 &   8.53 &  21.6  & MOIRCS & GY 287 & -\\
  8 &  16 27 38.95 &   -24 40 20.7 & 16.54 &  13.36 &  15.80 &  26.70 &  38.20 &  53.20 &  11.5  & MOIRCS & GY 312 & -\\
 9 &  16 27 41.80 &   -24 42 34.7 & 21.32 &  14.97 &   1.64 &   2.03 &   2.29 &   2.43 &  28.3  & MOIRCS & - & - \\
 10 &  16 27 48.24 &   -24 42 25.8 &  20.76 &  15.21 &   1.21 &   2.21 &   3.64 &  14.10 &  24.1  & MOIRCS & AOC J162748.24-244225.6 & -\\
\hline
 11 &  16 25 42.54 &   -24 43 51.1 & 15.61 &  14.03 &   0.99 &   0.66 &   0.47 &   0.39 &   3.0   & 2MASS & BKLT J162542-244350 & -\\
 12 &  16 26 04.58 &   -24 17 51.5 &  15.79 &  12.19 &  10.20 &   9.89 &   8.98 &  11.30 &  13.8   & 2MASS & BKLT J162604-241753 & -\\
 13 &  16 26 18.82 &   -24 26 10.5 & 14.84 &  12.14 &  10.50 &  11.50 &  13.90 &  20.50 &   9.0   & 2MASS & CRBR 2317.3-1925 & WGM99\\ 
 14 &  16 26 21.53 &   -24 26 01.0 & 12.57 &  10.92 &  25.00 &  20.30 &  19.30 &  19.10 &   3.4   & 2MASS & GY 5 & WGM99,N02 \\ 
15 &  16 26 21.90 &   -24 44 39.8 &  12.34 &  10.86 &  28.80 &  26.60 &  23.80 &  26.40 &   2.6   & 2MASS & GY 3 & N02,W05\\
16 &  16 26 23.81 &   -24 18 29.0 &  16.07 &  13.58 &   2.72 &   2.27 &   1.79 &   1.62 &   7.9   & 2MASS & CRBR 2322.3-1143 & CTK00\\ 
17 &  16 26 27.81 &   -24 26 41.8 & 14.26 &  12.09 &   9.52 &   8.74 &   7.78 &   7.38 &   6.2   & 2MASS & GY 37 & WGM99,W05 \\
 18 &  16 26 37.81 &   -24 39 03.2 & 14.98 &  12.85 &   4.47 &   3.92 &   3.35 &   3.21 &   6.0   & 2MASS & GY 80 & -\\
 19 &  16 26 51.28 &   -24 32 42.0 & 15.30 &  13.89 &   1.55 &   1.31 &   1.34 &   1.24 &   2.2   & 2MASS & GY 141 & LLR97,CTK00\\
 20 &  16 27 05.98 &   -24 28 36.3 & 16.64 &  12.97 &   5.43 &   5.16 &   5.32 &   5.37 &  14.1   & 2MASS & GY 202 & WGM99,LR99,CTK00\\
 21 &  16 27 26.22 &   -24 19 23.0 & 16.40 &  12.93 &   6.20 &   6.29 &   6.48 &   6.83 &  13.1   & 2MASS & BKLT J162726-241925 & -\\
 22 &  16 27 26.58 &   -24 25 54.4 & 13.00 &  11.84 &  10.30 &  10.20 &   9.78 &  12.00 &   0.9   & 2MASS & GY 264 & W05\\
 23 &  16 27 37.42 &   -24 17 54.9 &  14.15 &  11.95 &  11.20 &  10.20 &  10.20 &  11.70 &   6.4   & 2MASS & ISO-Oph 160 & N02\\
 24 &  16 27 38.95 &   -24 40 20.7 & 16.54 &  12.29 &  15.80 &  26.70 &  38.20 &  53.20 &  17.2   & 2MASS & ISO-Oph 165 & - \\ 
 25 &  16 27 40.84 &   -24 29 00.7 & 14.66 &  13.10 &   2.78 &   2.28 &   2.22 &   2.64 &   2.9   & 2MASS & GY 320 & -\\
 26 &  16 27 46.29 &   -24 31 41.2 & 13.83 &  11.32 &  17.60 &  15.70 &  14.80 &  18.60 &   8.0   & 2MASS & GY 350 & N02 \\ 
 27 &  16 28 17.74 &   -24 17 05.0 & 14.57 &  12.31 &   7.40 &   6.17 &   5.20 &   4.29 &   6.6   & 2MASS & BKLT J162817-241706 & -
\enddata
\tablenotetext{a}{References for papers spectroscopically confirming BD candidacy: WGM99: \citet{wilking99}, CTK00: \citet{cushing00}, N02: \citet{natta02}, W05: \citet{wilking05}. References for identifiers: GY: \citet{greene92}; CRBR: \citet{comeron93}; BKLT: \citet{barsony97}; ISO-Oph: \citet{bontemps01}; AOC: \citet{alves-de-oliveira08}}
\end{deluxetable}
\clearpage

For this sample of J\ks\ + Spitzer excess sources, estimates of the extinction are derived as described in Sect.\ \ref{ssec:ijk_candsel}. The visual extinction obtained for this sample is $A_V = 0 - 30$, and is given in Table \ref{tbl:subst_disks}.

\section{Multi-object spectroscopic follow-up of candidates}
\label{sec:mosspec}

\subsection{Observations and data reduction}
We used MOIRCS to carry out multi-object spectroscopy for 59 BD candidates selected from the i', J, \ks\ photometry (see Sect. \ref{sec:selection}), comprised of 58 new candidates and 1 previously known BD candidate, GY~84. The targets were covered in 3 fields, centered on the areas with highest density of candidates, indicated in Fig.\ \ref{fig:spatial}. The three fields are numbered by increasing RA as field 1, 2, and 3. Two of these candidates are located only 1.5\arcsec\ apart from each other in $\alpha$, so that they can be covered in one slit. In addition we observed one confirmed young M dwarf, GY84, published by \citet{natta06}, which falls into one of our fields.

Pre-imaging for the MOS masks in \ks-band was obtained in April 2009.
We used slits which are 0.9\arcsec\ wide and 9-12\arcsec~long. The spectroscopy run was carried out in two nights on May 30--31, 2009 using the grisms HK500, which covers the H- and \ks-band. The second half of the first night was lost due to technical problems, as a result the exposure time for field 1 is lower than intended. The total on-source time was 30, 140, and 150\,min for fields 1--3, split in shorter exposures of 5 or 10\,min. The typical seeing during these nights was 0.5\arcsec, as determined in the focusing procedure. The average airmass for the three fields is 1.8, 1.5, and 1.5.

Before and after the \rhooph\ masks, we observed A0 stars through one of the science masks for flux calibration and extinction correction; these exposures required de-focusing to prevent saturation. In total two A0 stars were observed in night 1 and seven in night 2, at various airmasses. For each mask, we obtained series of domeflats with lamp on and off for calibration purposes. Between the single exposures, the telescope was moved so that the targets shifted by 2.5\arcsec\ along the slit (nodding), to facilitate sky subtraction.

The MOS data was reduced following mostly the recipes outlined in Scholz et al. (2009). After subtracting the nodded exposures to remove the background, the images were divided by a normalized flatfield. Frames from the same nodding position were co-added using the median and a sigclip routine to remove cosmic ray events. This gives us two final frames for each chip and mask, from which the spectra were extracted using {\it apall} in IRAF. All candidates are well-detected. The wavelength solution was derived from the OH lines in the unreduced frames, for each object separately. The typical RMS of the fit was 2--3 \AA, well below the resolution. All spectra binned to 40 \AA~per pixel, corresponding to a resolution of 500 at 2.0 \micron. The two nodded spectra for the same object were co-added. The standard star spectra were treated in exactly the same way as the science frames.

After background subtraction the science frames still show 5-10\% residuals from the OH lines. The reason is that \rhooph\ was observed at relatively high airmasses, so that consecutive exposures differ in airmass by as much as 0.1. However, the extraction with {\it apall} includes a background fit perpendicular to the dispersion direction which reliably removes the residuals. We made sure that the final science spectra are not significantly affected by OH lines.

The corrections for the broad telluric absorption features between the near-infrared bands and for the instrument response was done in one step, using the A0 standard stars. The A0 standard star spectra were divided by a library spectrum of a A0 star \citep{pickles98}, obtained from the ESO website\footnote{http://www.eso.org/sci/facilities/paranal/instruments/isaac/tools/lib/}. The resulting calibration spectra show the combined effects of telluric absorption and response. Two out of the nine spectra differ significantly from the average; in one case probably due to effects of saturation, the other one has an usually large J--\ks\ colour of $0.2$ (obtained from 2MASS) which might indicate the presence of a dust shell. Both were not considered for the calibration. 

Most of the differences in the remaining 7 A0 spectra are caused by the variable airmass. For each wavelength bin we fit the fluxes linearly as a function of airmass and extrapolated to airmass of 1.0. While the extinction coefficients are consistent with the standard extinction law in the wavelength domain 1.45-1.8 \micron\ and 2.05-2.3 \micron, they vary strongly and irregularly in the water absorption features outside these bands, which cannot be reliably calibrated. The spectrum for airmass 1.0 was smoothed and fit with a second order polynomial over the H- and \ks-band. The result is our final calibration spectrum. All science spectra were extinction corrected using the coefficients derived from the standard spectra and the average airmass, and then divided by the calibration spectrum. 

\subsection{Spectral analysis}
\label{ssec:specana}

Our goal is to identify young sources with spectral types later than M5. As outlined in detail in \citet{scholz09}, these objects show a characteristic spectral shape in the near-infrared. In particular, their spectra have a clear peak in the H-band \citep{cushing05}. This feature is caused by water absorption on both sides of the H-band. The depth of these features depends strongly on effective temperature. While the H-band peak appears round in old field dwarfs, it is unambiguous triangular in young, low-gravity sources, likely due to the effects of Collisional Induced Absorption \citep{kirkpatrick06,witte09}. In addition, young brown dwarfs are expected to have flat or increasing \ks-band spectra with CO absorption bands at $\lambda >$ 2.3 \micron. 

\begin{figure}
\includegraphics[width=0.8\columnwidth]{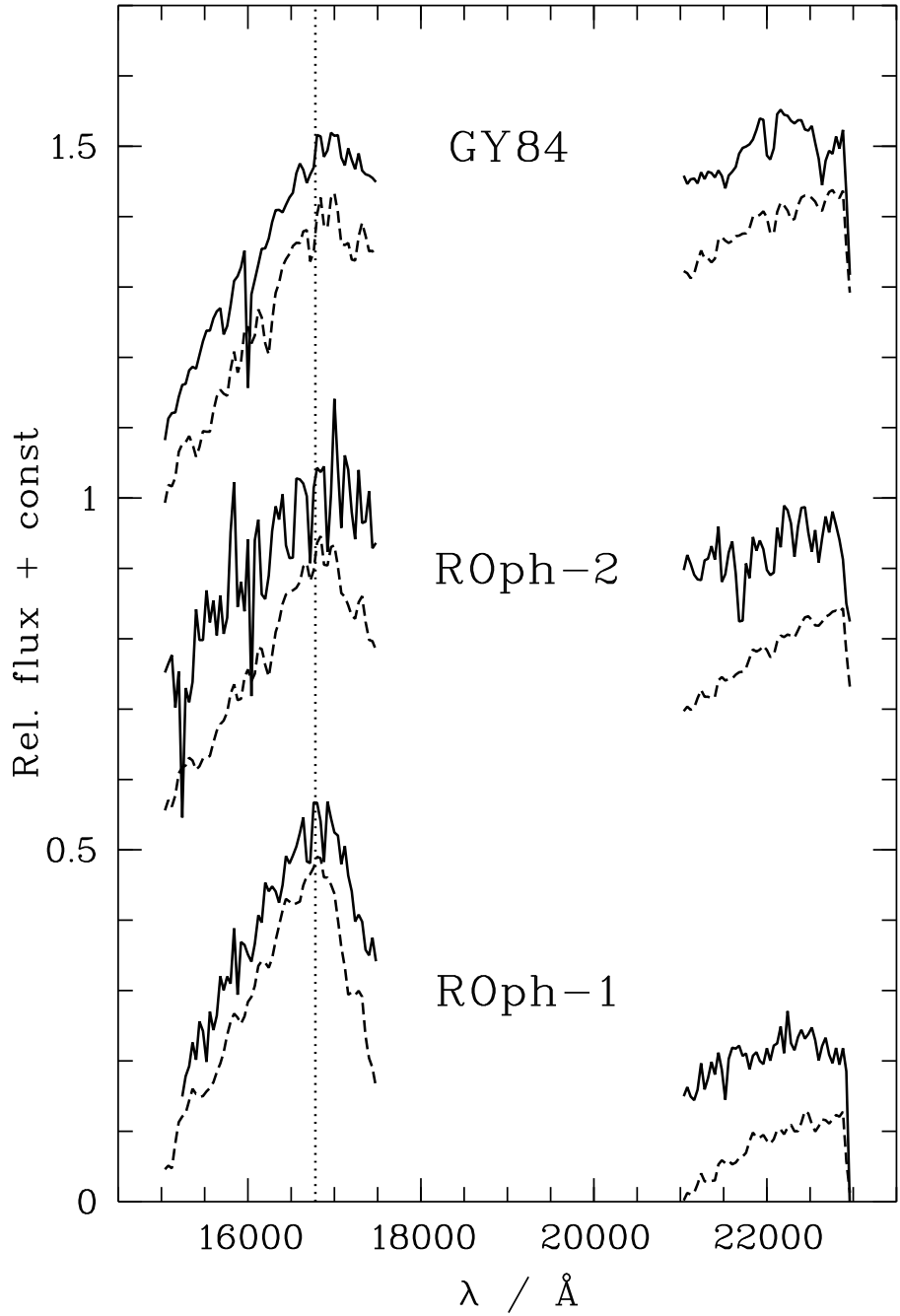}
\caption{MOIRCS spectra (solid) of candidate substellar objects in \rhooph, with best fit reddened model (dashed) with an \teff\ of 2500 K, 3100 K, and 3400 K; offsets were applied for clarity.}
\label{fig:spec}
\end{figure}

In a first step we visually inspect our spectra to search for the characteristic signature of young brown dwarfs. Three objects exhibit the features described above, including the object GY~84 from \citet{natta06}, these objects are listed in Table \ref{tbl:subst_spec} and their spectra are shown in Fig.\ \ref{fig:spec}. For six objects the signal-to-noise ratio is too low to conclusively determine their nature. The remaining 50 objects have smooth spectra with mostly decreasing \ks-band slope.

\begin{deluxetable}{lllllllll}
\tabletypesize{\scriptsize}
\tablecaption{Probable low mass and substellar mass members of \rhooph, with MOIRCS spectroscopy follow-up.\label{tbl:subst_spec}}
\tablewidth{0pt}
\tablehead{
\colhead{\#} & \colhead{RA (J2000)} & \colhead{Dec (J2000)} & \colhead{i' (mag)} & \colhead{J (mag)} & \colhead{\ks\ (mag)} & \colhead{\teff (K)} & \colhead{\av} & \colhead{Notes}}
\tablecolumns{9}
\startdata
1 & 16 26 56.33 & -24 42 37.8 & 21.24 & 17.68 & 15.53 & 2500 & 5 & SONYC-RhoOph-1 \\ 
2 & 16 26 57.36 & -24 42 18.8 & 22.53 & 18.94 & 15.73 & 3100 & 10 &  SONYC-RhoOph-2 \\
3 & 16 26 38.82 & -24 23 24.7 & 21.10 & 15.10 & sat & 3400 & 14 & GY~84
\enddata
\end{deluxetable}

In a second step we estimate the effective temperatures of the BD candidates by fitting model spectra from the DUSTY series \citep{allard01}, following the procedure outlined in \citet{scholz09}. We used the extinction as determined from the J--\ks\ colour (see Sect.\ \ref{ssec:jk_spitzer_candidates}). For GY84 the value of $A_V=14$ published by \citet{natta06} is used. The results are listed in Table \ref{tbl:subst_spec}. We find best-fitting effective temperatures of 2500\, and 3100\,K for our two candidates, here named SONYC-RhoOph-1, and SONYC-RhoOph-2, and a T$_{\mathrm{eff}}$ = 3400\,K for GY84, corresponding to spectral types M9, M5, and M3 respectively. SONYC-RhoOph-1 is confirmed as a newly found brown dwarf. Based on COND03 and DUSTY00 evolutionary tracks for 1Myr, it has a mass estimate of 0.015-0.02 \msun, marking it as one of the lowest mass objects in \rhooph. The temperatures we derive for GY 84 and SONYC-RhoOph-2 are too high to classify them as brown dwarfs. For comparison \citet{natta06} list T$_{\mathrm{eff}}=2900$ for GY84, derived by converting the J-band magnitude to luminosity and comparing with theoretical evolutionary tracks. The model spectrum with 2900\,K however clearly does not match the observed spectrum.

\begin{figure}
\includegraphics[width=0.8\columnwidth]{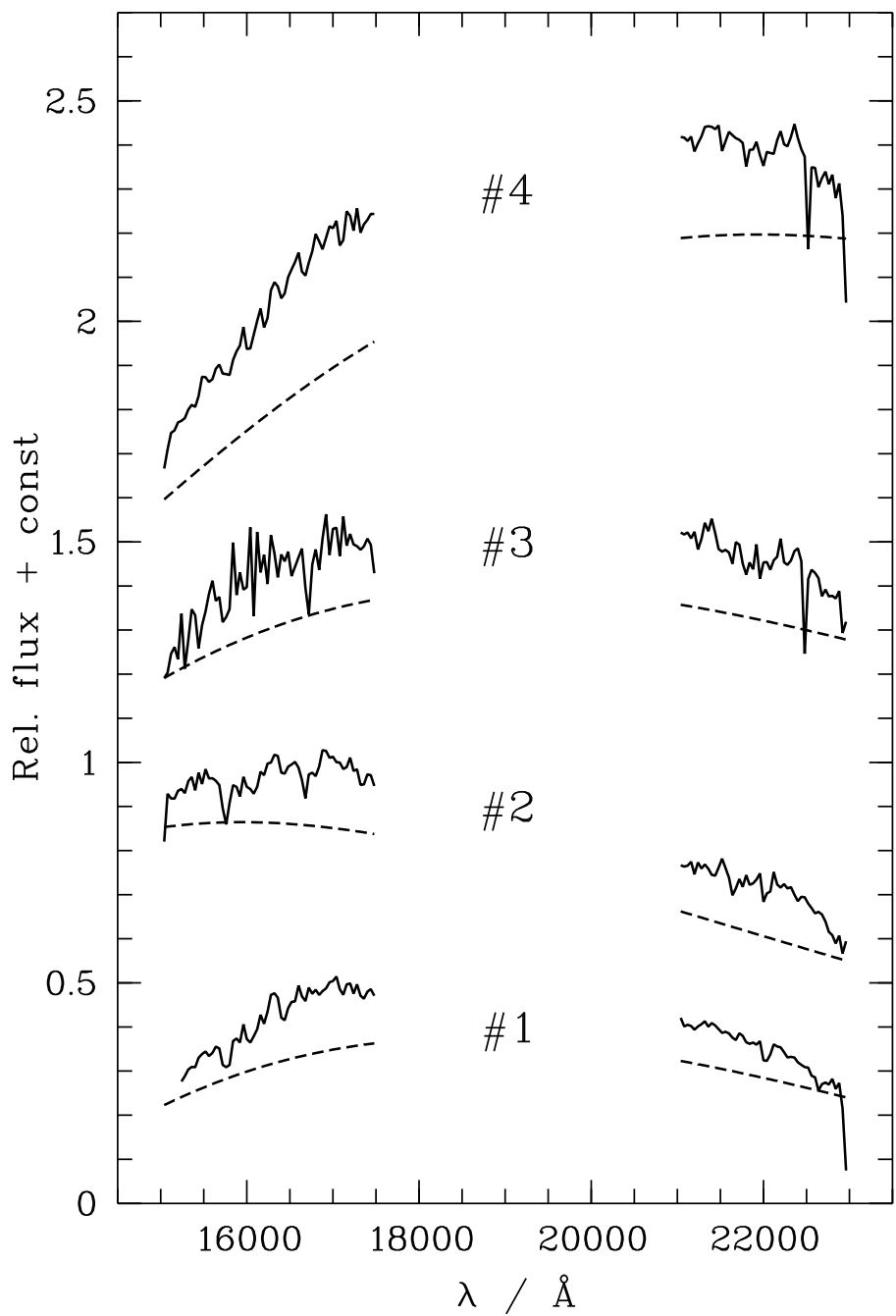}
\caption{Example MOIRCS spectra (solid) of sources in \rhooph\ which are not substellar candidates, with overplotted reddened blackbodies (dashed) with \teff\ ranging from 3000--4000 K; offsets were applied for clarity. The objects \#1 and \#4 in this plot are known in the literature as BKLT J162759-242912 and BKLT J162708-244227 \citep{barsony97}.}
\label{fig:otherspecs}
\end{figure}

The remaining sources, which are not brown dwarfs and have sufficient signal-to-noise ratio, exhibit smooth, featureless spectra, which makes it difficult to constrain their nature. 
The overwhelming majority of these objects are expected to be background late-type objects giant or dwarf stars. Some of them could be embedded young stellar objects with spectral types earlier than M, although we do not have any evidence for youth. We note that this sample contains 14 objects\footnote{BKLT J162640-242046, J162646-242155, J162654-244254, J162655-244242, J162656-244338, J162658-244342, J162703-244400, J162704-244422, J162708-244227, J162743-242822, J162745-242952, J162748-242542, J162750-242544, J162759-242912} 
from the BKLT catalog, a near-infrared survey by \citet{barsony97} without spectroscopic follow-up. Four examples of these featureless spectra are shown in Fig.\ \ref{fig:otherspecs}.

\subsection{Comparison with other studies}
\label{ssec:comparison}
The iJ\ks\ selected catalog of candidate brown dwarfs is compared with the list of 24 candidate BDs published in \citet[hereafter W08]{wilking08}. 21 objects fall within the spatial coverage of both the i'-band and the J+\ks-band imaging; 3 objects are not covered in i'- and/or J-band (GY141, Oph-160, Oph-193). Of the 21 covered, 12 are saturated (CRBR14, CRBR31, GY3, GY5, GY10, GY37, GY64, GY84, GY204, GY264, GY310, GY350, GY202), 3 are undetected in i-band (CRBR15, CRBR23, GY258), 1 is undetected in both i-band and J-band (GY31), and 1 is not properly detected in J-band due to strong distortion from a nearby bright source (see Sect.\ \ref{ssec:nirphot}). 3 of the 21 are recovered in the iJ\ks\ catalog, GY11, GY201, and GY325, indicated in Fig.\ \ref{fig:cmd_i_ij}. Of these 3, only 1 is recovered as a BD candidate (GY11) in the (i', i'-J) CMD selection (Sect.\ \ref{ssec:ijk_candsel}), while the other 2 (GY201, GY325) are blue-ward of the selection cutoff. 

We also compare our candidate BDs to the \citet{natta06} sample of objects with evidence of a circumstellar disk, and measurements of accretion. Three sources are covered in the iJ\ks\ catalog, ISO033 (=GY11 in W08), ISO165, and ISO169b, all three are recovered as brown dwarf candidates.

We compare our study with the recent paper by \citet{alves-de-oliveira10}, hereafter referred to as A10. Both studies use JH\ks\ near-infrared photometry; in this paper, i-band photometry is added for the entire survey area, while A10 added archive i-band and z-band photometry for a subset of their JH\ks\ sources. Both studies select candidate BDs from colour-magnitude and colour-colour diagrams based on predictions by evolutionary models. Two of the six new BDs found by A10 are covered in our iJ\ks\ survey area, but were not recovered in our iJ\ks\ catalog, due to saturation in J-band in one case and extended emission interference in i'-band in the other case. Our one new spectroscopically confirmed BD (RA 246.734 -24.711) appears to have been covered by A10 (based on their Fig. 4), but is not reported as a candidate Rho Oph member (A10, Table 4).

\section{Discussion: brown dwarf population in \rhooph}
\label{sec:discussion}

In the following section we use our survey results in combination with literature data to put limits on the total number of substellar objects in the \rhooph\ star forming region. We would like to preface this analysis with a cautionary note: \rhooph\ with its high and strongly varying extinction is a difficult field to search for brown dwarfs. All existing surveys have to be considered incomplete. The three key requirements to achieve a full census of BDs in this area are a) observations in multiple bands, b) combination of different survey strategies (e.g., broadband vs. narrowband surveys), and c) comprehensive spectroscopic follow-up. 

Our SONYC survey provides new constraints on the substellar population in \rhooph. We have used two different strategies. In a first step, we identify objects with iJ\ks\ photometry. As it turns out, this selection is heavily affected by background contamination and is not effective in the high extinction areas. It mostly covers the edges of the dense cloud core. In a second step we look for Spitzer/IRAC colour excess due to a disk in objects from our J\ks\ database. This sample should be dominated by young objects in \rhooph, but it is not clear what fraction is substellar. Since we do not require an optical counterpart, this selection also traces the dense cloud areas. Thus we provide two complementary approaches to identify brown dwarfs.

Our iJ\ks\ survey is certainly not complete for substellar objects. In the high extinction areas the i-band data is not deep enough (although deeper than all other previously published optical surveys in this region). In the low extinction areas, more massive brown dwarfs can be saturated in the J- and \ks-band, as evidenced by the significant number of literature BDs missing in our catalog (Sect.\ \ref{ssec:comparison}). Our sample identified from Spitzer colour excess is missing objects without disks. 

The previous surveys in the literature are subject to various biases as well. For example, the 20 objects observed by \citet{wilking99} have been selected from various near-infrared surveys which yield a much larger initial sample. Their colour criterion favours objects `near the surface of the cloud' \citep{wilking99} and will not find deeply embedded sources. The objects confirmed by \citet{natta02} as brown dwarfs in \rhooph\ have been selected from the ISOCAM survey by \citet{bontemps01} using a combined extinction (i.e. colour) and luminosity criterion. This approach will by definition not be able to find the objects without disks nor the ones with high extinctions. A similar argument can be made for the objects confirmed by \citet{jayawardhana06b} and \citet{allers07}, which were selected from Spitzer data. This simply illustrates that only the combination of multiple survey strategies can provide a more complete census.

One of the key features of our survey is extensive follow-up spectroscopy for our primary candidates from the iJ\ks\ database. Out of 504 objects, 309 were selected with colours consistent with substellar or planetary masses; for 58 of these candidates, and 1 known YSO, we obtained spectra. Only three of these turn out to be bona fide young stellar objects, one is a newly discovered young brown dwarf. Assuming that we selected the spectroscopy candidates in an unbiased way, we expect that the total catalog contains not more than a handful ($\sim 5$) of new substellar objects. 

The latter estimate is only a lower limit for the actual number of new BDs in our catalog. The spectroscopy fields were deliberately placed in regions with strong clustering. As the spectroscopy shows, most of the objects in these regions are background stars, i.e. the clustering reflects substructure in the extinction of the cloud. In areas with higher or lower extinction we expect to have less contamination by background objects because they are either not visible or do not pass our colour criterion.
Taking this into account, we make a second estimate of the total number of BDs in the iJ\ks\ catalog based on the success rate and spatial coverage of the follow-up spectroscopy while we assume a spatially homogeneous population of brown dwarfs. Our three MOS fields cover 72 sq. arcmin out of $\sim 616$ sq. arcmin (0.171 sq. deg.) in the total i+J\ks\ survey ($\sim 12$\%). We obtained spectra for 59 out of the 159 iJ\ks\ candidates ($\sim 37$\%) located within the area of the MOS fields and find one bona fide brown dwarf. Scaling to the total area of our survey yields an estimate of 24 brown dwarfs. We note this estimate is taken as an order of magnitude estimate, since it has a large uncertainty (1 $\sigma$ confidence interval of 5--74, assuming binomial distribution), and moderate clustering in the YSO population \citep{bontemps01} may further increase the estimate. We conclude that despite the low yield of our spectroscopy campaign our database might still contain a substantial number of new brown dwarfs. Spectroscopy covering the full field is required to clarify this.

We demonstrate that the available Spitzer data contains more objects than previously known that might be BDs in \rhooph. By combining our deep J\ks\ data with the Spitzer C2D catalog, we select 10 candidates with mid-infrared colour excess and near-infrared colours indicative for a substellar mass, of which 1 is a previously spectroscopically confirmed BD. In addition, 17 objects were found with 2MASS J\ks\ data in combination with Spitzer data, using the same approach, of which 10 objects were previously spectroscopically confirmed BDs. The 17 new candidates exhibit a wide range of visual extinctions from 0 to 20\,mag. In particular the high extinction objects constitute a parameter regime not covered sufficiently in previous surveys. Again, this sample requires full spectroscopy follow-up for a more detailed characterisation.

In Table \ref{tbl:rhoophbds} we provide a list of the spectroscopically confirmed brown dwarfs in the literature, including our new object, restricted to our i'-band survey area (246.415 $\le$ RA $\le$ 247.085, -24.77 $\le$ DEC $\le$ -24.275). The main criteria for an object to appear in this list are a) spectral type later than M6 or alternatively effective temperature below 3000\,K and b) evidence for youth (e.g., disk excess, spectral features indicative of low gravity). We include sources with multiple spectroscopically determined spectral type estimates, if at least one is M6 or later. To date, 21 objects fulfill these criteria. As outlined above in detail, this number is almost certainly a lower limit to the full number of BDs in \rhooph.  The table should be seen as a starting point in establishing a more complete census. We did not include the three brown dwarfs from \citet{allers07} as those are located outside our survey area. We note the inclusion of the spectroscopically confirmed early T-dwarf found by \citet{marsh10a}, for which evidence for youth is found in low gravity features. This may be the lowest mass object identified in \rhooph\ so far, although its cluster membership is still in question as \citet{alves-de-oliveira10} note that they find a different \ks\ magnitude than \citet{marsh10a}, based on which the distance estimate would place the object at 137 -- 217 pc, behind the \rhooph\ cloud.
Finally, we note that 16 of the 21 objects in Table \ref{tbl:rhoophbds} are in agreement with \citet[Table 7]{alves-de-oliveira10}. We do not list the 5 of their BDs outside our survey area: ISO-Oph 193, CFHTWIR-Oph 4, 47, 57, and 106. We do list 5 BD candidates that they do not : the newly found SONYC-RhoOph-1 and MARSH10-4450, as well as the 3 borderline cases GY 37, GY 59, and GY 84, that have multiple conflicting spectral type estimates in the literature.

\begin{deluxetable}{llllllll}
\rotate
\tabletypesize{\scriptsize}
\tablecaption{Spectroscopically confirmed Brown Dwarfs within i'-band survey area of \rhooph.\label{tbl:rhoophbds}}
\tablewidth{0pt}
\tablehead{
\colhead{Name(s)} & \colhead{RA (J2000)} & \colhead{Dec (J2000)} & \colhead{Indicator(s) of youth} & \colhead{SpT} & \colhead{References\tablenotemark{a}}}
\tablecolumns{6}
\startdata
CRBR 2317.3-1925 / CRBR 14 & 16 26 18.82 & -24 26 10.5 & CO bands+\av, mid-IR excess & M7.5,M5.5,M7 & WGM99,LR99,N02 \\ 
GY 5 & 16 26 21.54 & -24 26 01.0 & CaH, TiO, CO bands+\av, mid-IR exc. & M5.5,M7,M6 & W05,WGM99,N02 \\
GY 3 & 16 26 22.05 & -24 44 37.5 & CaH, TiO, mid-IR excess & M8,M7.5 & W05,N02 \\
GY 10 & 16 26 22.17 & -24 23 54.4 & CO bands+\av & M8.5,M6.5 & WGM99,LR99 \\
GY 11 & 16 26 22.28 & -24 24 09.3 & H$_2$O abs., CO bands+\av & M6.1,M6.5,M8.5 & CTK00,WGM99,N02 \\
CRBR 2322.3-1143 / CRBR 31 & 16 26 23.78 & -24 18 31.4 & H$_2$O abs. & M6.7 & CTK00  \\ 
GY 37 & 16 26 27.83 & -24 26 42.6 & CaH, TiO & M5,M6 & W05,WGM99 \\
GY 59 & 16 26 31.37 & -24 25 30.3 & CO bands+\av & M3.75,M6,M5 & W05,WGM99,LR99 \\
GY 64 & 16 26 32.56 & -24 26 36.9 & H$_2$O abs., CO bands+\av & M7.0,M8 & CTK00,WGM99 \\
GY 84 & 16 26 38.80 & -24 23 22.7 & CO bands+\av, H$_2$O abs. & M6,M3\tablenotemark{b} & WGM99, this work \\
CFHTWIR-Oph 34 & 16 26 39.92 & -24 22 33.6 & H$_2$O absorption & M8.25 & A10\\
GY 141 & 16 26 51.42 & -24 32 42.7 & H$_{\alpha}$, H$_2$O abs. & M8.5,M8.0 & LLR97,CTK00 \\ 
SONYC-RhoOph-1 & 16 26 56.33 & -24 42 37.8 & H$_2$O absorption & M9\tablenotemark{b} & this work \\
GY 202 & 16 27 06.00 & -24 28 37.3 & H$_2$O abs., CO bands+\av & M5.7,M7,M6.5 & CTK00,WGM99,LR99 \\
GY 204 & 16 27 06.58 & -24 41 47.9 & CaH, TiO, mid-IR excess & M5.5,M6 & W05,N02 \\
MARSH10-4450\tablenotemark{c} & 16 27 25.35 & -24 25 37.5 & CH$_4$, H$_2$O absorption & early T & M10\\
GY 264 & 16 27 26.58 & -24 25 55.1 & CaH, TiO & M8 & W05 \\
ISO-Oph 160 & 16 27 37.30 & -24 17 56.4 & mid-IR excess & M6 & N02 \\ 
GY 310 & 16 27 38.67 & -24 38 38.2 & CO bands+\av & M8.5,M7,M6 & WGM99,LR99,N02 \\
CFHTWIR-Oph 96 & 16 27 40.84 & -24 29 00.8 & H$_2$O absorption & M8.25 & A10\\
GY 350 & 16 27 46.36 & -24 31 41.6 & disk excess & M6 & N02
\enddata
\tablenotetext{a}{LLR97: \citet{luhman97}, WGM99: \citet{wilking99}, LR99: \citet{luhman99}, CTK00: \citet{cushing00}, N02: \citet{natta02}, W05: \citet{wilking05}, W08: \citet{wilking08}, A10: \citet{alves-de-oliveira10}, M10: \citet{marsh10a}. References for identifiers: GY: \citet{greene92}, CRBR: \citet{comeron93}, ISO-Oph: \citet{bontemps01}}
\tablenotetext{b}{Spectral type estimate based on \teff\ -- SpT scale from Table 8 of \citet{luhman03b}}
\tablenotetext{c}{Provisional name chosen in this paper; no entry for this source exists in SIMBAD}
\end{deluxetable}
\clearpage

Our sample of candidates identified from Spitzer excess allows us to estimate an upper limit for the number of missing BDs in \rhooph. Our MOIRCS J\ks\ survey area covers 19 of the 21 spectroscopically confirmed brown dwarfs from Table \ref{tbl:rhoophbds}. In the same area are an additional 16 new BD candidates with disks (see Table \ref{tbl:subst_disks}). Assuming a disk fraction of $\sim 67\%$ \citep{jayawardhana03b}, the upper limit for BD candidates in this area is $\sim 16 \times 1.5$ = 24, and subsequently, an upper limit on total number of BDs would be 24 + 19 = 43, roughly twice the number already confirmed with spectroscopy within our J\ks-survey area. In other words, the current census of confirmed BDs might still miss a few tens of substellar objects.

It is premature to derive the substellar IMF in \rhooph. However, based on the currently available data we can estimate the number of low-mass stars ($0.1<M<1.0\,\msun$) in relation to the number of BDs ($M<0.1\,\msun$), a ratio that has been derived for a number of star forming regions \citep{andersen08,scholz09}. 

\citet{bontemps01} provide constraints on the stellar mass function in this region, complete down to below $0.1\,\msun$.  From their Fig.\ 8 we infer a lower limit of $\sim 80$ stars in the given mass range. Taking into account an estimated binarity of $\sim 29\%$ \citep{ratzka05} would increase this number to about 103. \citet{wilking08} summarized a list of 316 L1688 association members with at least a \ks-band 2MASS detection, as well as one or more signs of association membership, and 169 out of 316 sources have an optical and/or IR spectral type estimate. Of these 169, 19 association members are Brown Dwarfs with one or more spectral type estimations of M6 or later, while 140 are low mass stars with spectral type estimates between G2 and M6 ($0.1<M<1.0\,\msun$). Combining mid- and near-infrared as well as X-ray data, these are the most complete estimates for the stellar population in \rhooph\ to date. Making the reasonable assumption that the stellar sample is more complete than the brown dwarf sample, and adopting a range of 103--140 in number of stellar sources, and 19 brown dwarfs, then an upper limit for the ratio low-mass stars to BDs is 5--7. Depending on how many BDs the current census is missing, this ratio might be as low as $\sim$ 3. As it stands now, the observed ratio is fully in line with the literature values for most other young clusters (3.3-8.5, \citealt{andersen08}), although significantly higher than the value we have obtained for NGC1333 (1.5, \citealt{scholz09}). Note that the values quoted by \citet{andersen08} should be treated as upper limits as well, since they are limited to BDs with $M>0.03\,\msun$. Thus, there is no reason to believe that \rhooph\ has a substellar IMF significantly different from other star forming regions.

\section{Conclusions}
\label{sec:conclusions}
\begin{itemize}
\item Large-scale, optical+near-infrared imaging surveys with Suprime-Cam and MOIRCS have been used to create a catalog of sources, reaching completeness limits of 24.2 in iÕ-band, 20.6 in J-band, and 17.8 in \ks-band. The spatial coverage of our i'+J\ks\ survey is 0.171 sq.\ deg.\ and covers the L1688 core in i'-, J-, and \ks-band.

\item In terms of object masses for members of \rhooph, the survey completeness limits correspond in the i'-band to mass limits of 0.004 -- 0.1 \msun, and in the J-band to mass limits of 0.001--0.007 \msun, for extinction ranges of \av\ = 5--15, based on the COND03 and DUSTY00 evolutionary tracks. 

\item From the optical + near-infrared photometry, 309 objects were selected as candidate substellar and planetary-mass cluster members. 58 of these objects, and 1 additional previously known BD candidate, were targeted for follow-up spectroscopy. Based on multi-object spectroscopy in H-band and \ks-band, using the water absorption features in the H-band, 1 of the 58 new candidates was confirmed as a substellar mass object with T$_{\mathrm{eff}}$ = 2500 K.

\item Based on literature and this survey, we identify a sample of 21 spectroscopically confirmed young Brown Dwarf members within our survey area of \rhooph. However, this sample, based on current existing surveys, is likely highly incomplete, due to the variable and high extinction in \rhooph. 

\item From MOIRCS, 2MASS, and Spitzer photometry a sample of 27 sources with mid-infrared colour excess and near-infrared colours indicative for substellar mass sources with disks are identified. Of these, 11 are previously spectroscopically confirmed brown dwarfs, while 16 are newly identified candidates.

\item Based on present day surveys of the stellar and brown dwarf populations, the ratio of substellar to stellar sources in \rhooph\ is derived to have an upper limit of 5--7. This is in line with other nearby young star forming regions.
\end{itemize}

\acknowledgments
The authors would like to thank Kentaro Aoki for supporting the spectroscopy run with Subaru/MOIRCS, Laura Fissel for discussion on the image reduction, and an anonymous referee for useful suggestions toward improving the paper. AS acknowledges travel support through the UK/PATT grant SPA0-YST003. RJ also acknowledges support from a Royal Netherlands Academy of Arts
and Sciences (KNAW) visiting professorship. This publication makes use of data products from the Two Micron All Sky Survey, which is a joint project of the University of Massachusetts and the Infrared Processing and Analysis Center/California Institute of Technology, funded by the National Aeronautics and Space Administration and the National Science Foundation.  This research makes use of the SIMBAD database, operated at CDS, Strasbourg, France.

\bibliographystyle{apj}
\bibliography{fullbib}

\end{document}